\documentclass[11pt]{article}
\usepackage{amsmath,amssymb,amscd,amsthm}
\usepackage{mathrsfs}
\usepackage[dvipdfmx]{graphicx}
\usepackage{cite}
\usepackage{color}
\usepackage{enumitem}
\usepackage{ascmac}
\usepackage{fullpage}
\usepackage{microtype}

\definecolor{darkblue}{rgb}{0.1,0.1,.7}
\usepackage[breaklinks=true,linktocpage=true,
colorlinks=true,urlcolor=darkblue,linkcolor=darkblue,
citecolor=darkblue,pdfpagelabels=true,hypertexnames=true,
plainpages=false,naturalnames=false,]{hyperref}

\numberwithin{equation}{section}
\newcommand{\Ga}{{\Gamma}}
\newcommand{\De}{{\Delta}}

\newcommand{\bt}{{\beta}}
\newcommand{\ga}{{\gamma}}
\newcommand{\de}{{\delta}}
\newcommand{\ep}{{\epsilon}}
\newcommand{\vep}{{\varepsilon}}
\newcommand{\zt}{{\zeta}}
\newcommand{\te}{{\theta}}

\newcommand{\ka}{{\kappa}}
\newcommand{\lm}{{\lambda}}

\newcommand{\sig}{{\sigma}}
\newcommand{\vphi}{{\varphi}}

\newcommand{\cA}{{\mathcal{A}}}

\newcommand{\cD}{{\mathcal{D}}}

\newcommand{\cN}{{\mathcal{N}}}
\newcommand{\cO}{{\mathcal{O}}}

\newcommand{\cQ}{{\mathcal{Q}}}

\newcommand{\frh}{{\mathfrak{h}}}

\newcommand{\mcd}{{\kern-0.1pt\cdot\kern-0.1pt}}
\newcommand{\nab}{{\nabla}}
\newcommand{\nn}{{\nonumber}}

\newcommand{\ol}{\overline}
\newcommand{\pd}{{\partial}}

\newcommand{\wt}{\widetilde}

\newcommand{\tr}{{\mathop{\mathrm{tr}}\nolimits}}

\def\Re{\mathop{\mathrm{Re}}\nolimits}

\newcommand{\bra}{{\langle}}
\newcommand{\ket}{{\rangle}}

\newcommand{\bbra}{{\langle\kern-2.5pt\langle}}
\newcommand{\kket}{{\rangle\kern-2.5pt\rangle}}
\newcommand{\Bbra}{{\Big\langle\kern-3.5pt\Big\langle}}
\newcommand{\Kket}{{\Big\rangle\kern-3.5pt\Big\rangle}}

\begin{document}

\thispagestyle{empty}

\noindent
\vskip2.0cm

\begin{center}

{\Large\bf Hybrid inflation from supersymmetry breaking}

\vglue.3in

\hspace{-2pt}Yermek Aldabergenov,${}^{a,b,}$\footnote{ayermek@fudan.edu.cn} Ignatios Antoniadis,${}^{c,d,}$\footnote{antoniad@lpthe.jussieu.fr} Auttakit Chatrabhuti,${}^{c,}$\footnote{auttakit.c@chula.ac.th} Hiroshi Isono${}^{c,}$\footnote{hiroshi.isono81@gmail.com}
\vglue.1in

${}^a$~{\it Department of Physics, Fudan University, 220 Handan Road, Shanghai 200433, China}\\
${}^b$~{\it Department of Theoretical and Nuclear Physics, 
Al-Farabi Kazakh National University,\\ 71 Al-Farabi Ave., Almaty 050040, Kazakhstan}\\
${}^c$~{\it High Energy Physics Research Unit, Faculty of Science, Chulalongkorn University, Phayathai Road, Pathumwan, Bangkok 10330, Thailand}\\
${}^d$~{\it Laboratoire de Physique Th\'eorique et Hautes Energies (LPTHE), Sorbonne Universit\'e,\\ CNRS, 4 Place Jussieu, 75005 Paris, France}\\
\vglue.1in

\end{center}

\vglue.3in

\begin{center}
{\Large\bf Abstract}
\vglue.2in
\end{center}

We extend a recently proposed framework, dubbed inflation by supersymmetry breaking, to hybrid inflation by introducing a waterfall field that allows to decouple the supersymmetry breaking scale in the observable sector from the inflation scale, while keeping intact the inflation sector and its successful predictions: naturally small slow-roll parameters, small field initial conditions and absence of the pseudo-scalar companion of the inflaton, in terms of one free parameter which is the first order correction to the inflaton K\"ahler potential. During inflation, supersymmetry is spontaneously broken with the inflaton being the superpartner of the goldstino, together with a massive vector that gauges the R-symmetry. Inflation arises around the maximum of the scalar potential at the origin where R-symmetry is unbroken. Moreover, a nearby minimum with tuneable vacuum energy can be accommodated by introducing a second order correction to the K\"ahler potential. The inflaton sector can also play the role of the supersymmetry breaking `hidden' sector when coupled to the (supersymmetric) Standard Model, predicting a superheavy superparticle spectrum near the inflation scale. Here we show that the introduction of a waterfall field provides a natural way to end inflation and allows for a scale separation between supersymmetry breaking and inflation. Moreover, the study of the global vacuum describing low energy Standard Model physics can be done in a perturbative way within a region of the parameter space of the model. 

\newpage

\tableofcontents

\setcounter{footnote}{0}

\section{Introduction}

In past works~\cite{Antoniadis:2017gjr, Antoniadis:2018cpq, Antoniadis:2019dpm}, a framework of intimate connection between supersymmetry and inflation was introduced, dubbed inflation by supersymmetry breaking, relating two theoretical proposals motivated, correspondingly, by particle physics and cosmology. The basic idea is to identify the inflaton with the superpartner of the goldstino,\footnote{For past related literature, see e.g.~\cite{Jeannerot:2000sv,Jeannerot:2002wt,sgoldstino1,sgoldstino2,sgoldstino3,sgoldstino4,sgoldstino5,susybreaking1,susybreaking2,susybreaking3,susybreaking4,susybreaking5}.} charged under a gauged R-symmetry with the corresponding $U(1)$ field becoming massive by absorbing the (complex) inflaton phase which is the R-goldstone boson. The superpotential is forced to be linear in the inflaton superfield $X$ by symmetry, while the K\"ahler potential can be expanded around its canonical form in powers of $X{\bar X}$ with small coefficients representing quantum corrections of the underlying supergravity theory. For a positive sign of the first order correction, neglecting D-term contributions suppressed by a small R-gauge coupling, the scalar potential has a maximum that can realise hilltop inflation with a spectral index of primordial scalar perturbations determined by this correction. Moreover, a second order correction allows for accommodating a nearby minimum with tuneable vacuum energy. The (supersymmetric) Standard Model (SM) can be coupled in a straightforward way with the inflaton part playing also the role of the supersymmetry breaking sector, while the gauge R-symmetry may contain the usual R-parity as a subgroup. The model is very predictive but leads to a superheavy spectrum of superparticles near the inflation scale.

In this work we address the problem of scale separation, aiming to decouple the scale of inflation from the one of low energy supersymmetry breaking but staying always within the successful framework of inflation by supersymmetry breaking. To this end,
we generalise the above framework to hybrid inflation~\cite{Linde:1991km,Linde:1993cn}, by introducing a waterfall direction in the scalar potential that opens up at a critical point at the end of inflation along which the waterfall field falls rapidly to a deep global minimum while the inflaton is displaced slightly. The waterfall field is neutral under the R-symmetry and its superpotential can be adjusted so that the global minimum stays infinitesimally close to a supersymmetric one which consists of a valley around a flat direction for the inflaton field. This allows for a perturbative treatment of the whole dynamics around the origin for the inflaton field and around the supersymmetric minimum for the waterfall with the supersymmetry breaking scale being a free parameter independent of the inflation scale. Moreover, without affecting the inflationary predictions of the model, the presence of the waterfall direction provides a way to end inflation efficiently and tune the vacuum energy of the global minimum at a value infinitesimally close to zero. The cancellation occurs between the negative F-term and positive D-term contributions to the scalar potential.\footnote{Previous works on hybrid inflation in the supersymmetric framework can be found for example in~\cite{Dvali:1994ms}. See also~\cite{hybrid1,Ahmed:2022dhc,Pallis:2024mip} and references therein for recent developments.}

The outline of the paper is the following. In Section~2, we present the model we study in this work containing the inflation (dubbed hidden) and the observable SM sector. The inflation sector contains the inflaton chiral superfield charged under an abelian vector field that gauges the $U(1)_R$ symmetry, as well as the waterfall chiral superfield which is neutral under $U(1)_R$. We give the K\"ahler potential and superpotential that determine the $N=1$ supergravity effective action, assuming a constant (field independent) R-gauge kinetic function. We describe the setup of inflation and recall its predictions. In Section~3, we analyse the vacuum structure of the model, first without and then with the D-term potential contribution. The analysis is perturbative around the origin of the inflaton direction and to the leading order in the supersymmetry breaking scale which can be parametrically small compared to the inflation scale. In Section~4, we impose theoretical and observational constraints and determine the allowed parameter region. In Section~5, we discuss the supersymmetry breaking and the particle spectrum in both hidden and observable sectors of the theory. We end with our conclusions in Section~6. Finally, there are two appendices with technical details on the computation of the parameter space and of fermion masses.

\section{Model}

\subsection{Setup}
\label{subsec:Model}

\subsubsection{Inflaton sector}
We work with four-dimensional $\cN=1$ supergravity theories which contain two chiral multiplets and an abelian vector multiplet that is associated with a gauged $U(1)_R$ transformation. We denote the two chiral superfields by $X,\phi$, and suppose that $X$ transforms as $X\mapsto e^{-iq\te}X$ but $\phi$ is neutral under the $U(1)_R$. Our model is defined by the following K\"ahler potential, superpotential and gauge kinetic function,
\begin{align}
K&=X\ol X+\ka^2A(X\ol X)^2+\ka^4B(X\ol X)^3+[1-\ka^2zX\ol X+\ka^4\ga(X\ol X)^2]\phi\bar\phi, \label{kahler} \\
W&=\ka^{-2}fXw(\phi), \qquad w(\phi)=1+\ka^2\mu\phi^2+\ka^4\lm\phi^4, \label{super} \\
h&=1+\bt_R\ln(\ka X), \label{gk}
\end{align}
where $\ka^{-1}=2.4\times10^{15}$\,TeV is the reduced Planck mass. The superfields $X,\phi$ have mass dimension 1, while the parameters $A,B,z,\ga,f,\mu,\lm,\bt_R$ are dimensionless. In the rest of this paper, we set $\ka=1$. The parameters $A,B,z,\ga,f,\mu$ are (or can be chosen to be) real numbers, while $\lm$ -- although complex in general -- is assumed to be also real for simplicity. The superpotential has $U(1)_R$ charge $q$. The logarithmic correction in $h$ is introduced for the anomaly cancellation involving the gauged $U(1)_R$ (Green-Schwarz mechanism)~\cite{Freedman:2005up,Elvang:2006jk,Antoniadis:2014iea}. The coefficient $\bt_R$ is proportional to $q^2$. We will neglect this logarithmic term as the charge squared $q$ will be supposed to be tiny, except when we discuss the $U(1)_R$ gaugino mass which is determined by the logarithmic correction.

We parametrise the scalar fields $X,\phi$, which are the lowest components of the superfields $X,\phi$, respectively, as 
\begin{align}
X=\rho e^{i\upsilon}, \quad \phi=\phi_R+i\vphi,
\end{align}
where $\rho,\upsilon,\phi_R,\vphi$ are real fields. When $\rho\neq 0$, the phase $\upsilon$ is absorbed to make the $U(1)_R$ gauge field massive. We will identify $\rho$ with the inflaton.

The scalar potential is the sum of the $F$- and $D$-term potentials $V=V_F+V_D$, where each one is given by,\footnote{The covariant derivatives are defined by $D_iW=W_i+K_iW$, where subscripts denote differentiation with respect to the corresponding field. The K\"ahler metric is $g_{i\bar j}=\pd_i\pd_{\bar j}K$ and the inverse K\"ahler metric is defined by $g^{\bar ij}g_{j\bar k}=\de^{\bar i}{}_{\bar k}$ and $g_{i\bar j}g^{\bar jk}=\de_i{}^k$. The indices $i,j,k$ take on the chiral superfields $X$ and $\phi$.}
\begin{align}
&V_F=e^K(D_{\bar i}\ol WK^{\bar ij}D_jW-3W\ol W), \label{VF-genfml} \\
&V_D=\frac{1}{2}q^2\Re(\frh)\cD^2, \qquad \cD:=XK_X+XW_X/W=1+XK_X, \label{VD-genfml}
\end{align}
The potential is invariant under $\phi\to\bar\phi$ plus $X\to\ol X$ because it is real. It is also invariant under $X\to\ol X$ because it is independent of the phase of $X$. Therefore, $\phi\to\bar\phi$, equivalent to $\vphi\to-\vphi$, is a symmetry of the potential. Furthermore, the potential is invariant under $\phi\to-\phi$ alone because this keeps $K$ and $W$ invariant and the expression \eqref{VF-genfml} contains even numbers of $\phi$ and $\bar\phi$. Combining them yields the invariance under $\phi_R\to-\phi_R$. It is therefore enough to consider the region $\phi_R\geq 0$ and $\vphi\geq 0$.

\subsubsection{Coupling with supersymmetric Standard Model sector}
The model given above can be coupled with the supersymmetric Standard Model (MSSM). The inflaton sector then plays a role of supersymmetry breaking sector. Following~\cite{Aldabergenov:2021uye}, we consider two types of the MSSM superpotential: one in which the MSSM superfields are all neutral under $U(1)_R$, and the other in which some of the MSSM superfields are charged under $U(1)_R$ in such a manner that the MSSM superpotential has $U(1)_R$ charge $q$. Concretely, we will consider the following two models~\cite{Aldabergenov:2021uye}:
\begin{align}
\mbox{Model 1:} &~~
\hat K=K+\sum\zt\bar\zt, \quad
\hat W_1=X[fw(\phi)+\mu_HH_uH_d+y_u\bar u\cQ H_u-y_d\bar d\cQ H_d-y_e\bar eLH_d], \\
\mbox{Model 2:} &~~
\hat K=K+\sum\zt\bar\zt, \quad
\hat W_2=X[fw(\phi)+\mu_HH_uH_d]+y_u\bar u\cQ H_u-y_d\bar d\cQ H_d-y_e\bar eLH_d, \label{model2-W}
\end{align}
where $\zt$ collectively denotes the MSSM matter chiral superfields $H_u,H_d,\bar u,\bar d,\cQ,\bar e,L$, and the sum is over $\zt$. The vacuum expectation value of $\zt$ is zero. In Model 1, the MSSM superfields are neutral under $U(1)_R$. In Model 2, $H_u$ and $H_d$ are neutral while the other MSSM superfields have R-charge $q/2$, so that $\hat W_2$ has R-charge $q$. Under this charge assignment, the $U(1)_R$ group contains the usual R-parity as its subgroup.

\subsection{Scenario}
Our scenario is inflation followed by waterfall (see Figure~\ref{infwf}):
\begin{figure}
\centering
  \includegraphics[width=0.8\linewidth]{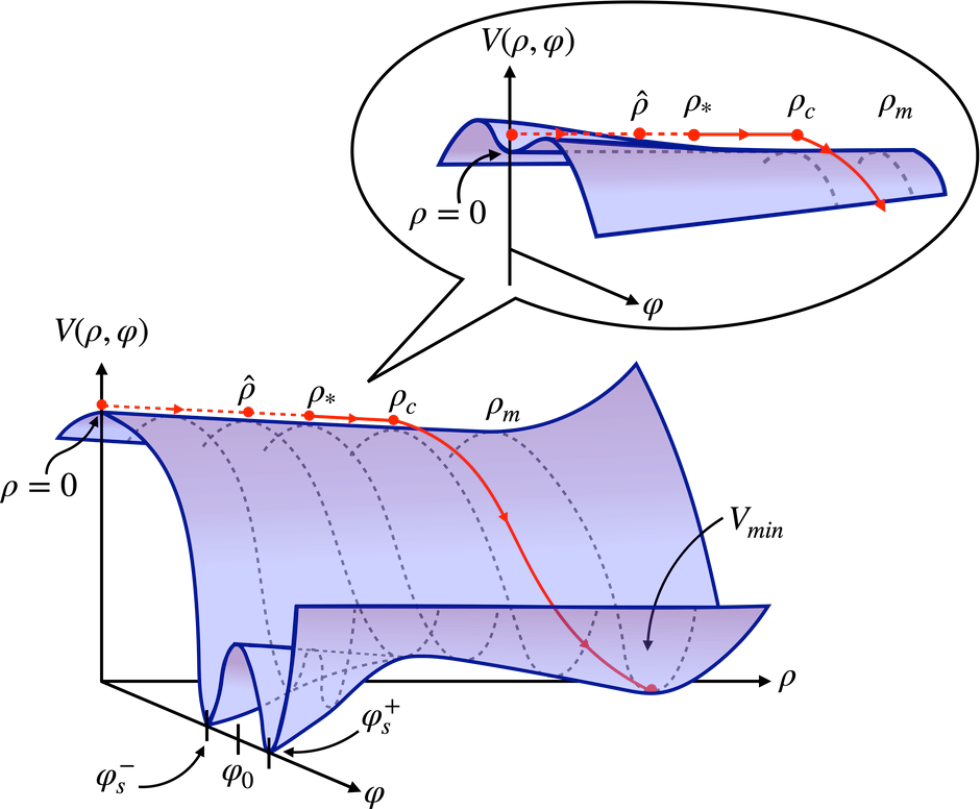}
\caption{Inflation followed by waterfall. The figure in the bubble magnifies the detailed shape of the potential during the inflation, which is a shallow valley. $\vphi_s^\pm$ and $\vphi_0$ will be introduced explicitly in Section~\ref{subsec:Vrho0}. The real part $\phi_R$ is suppressed because the potential is stable in this direction and the trajectory always runs with $\phi_R=0$.}\label{infwf}
\end{figure}
First, the inflaton is located near the origin $(\rho,\phi)=(0,0)$, and then rolls down slowly. During this slow-roll, the potential is stable in $\phi$, forming a valley along the $\rho$-axis. In the meantime the inflaton reaches a ``critical'' point $\rho_c$, where the potential becomes unstable in some direction in $\phi$. The inflaton then deviates from the $\rho$ axis, falls in this unstable direction (waterfall field $\phi$ is turned on), and reaches a vacuum at $(\rho_m,\phi_m)$. We assume that the inflaton still moves in the $\rho$ direction after $\rho_c$ and hence $\rho_c$ and $\rho_m$ satisfy
\begin{align}\label{rhoc<rhom<1}
\rho_c<\rho_m<1,
\end{align}
where the last inequality guarantees perturbative treatment.

Let us analyse the structure near the origin $\rho=\phi=0$. The mass term for $\phi$ at each $\rho$ is
\begin{align}
\frac{1}{2}m_R^2(\rho^2)\phi_R^2+\frac{1}{2}m_I^2(\rho^2)\vphi^2.
\end{align}
The mass functions at the origin $\rho=0$ are given by
\begin{align}
m_R^2(0)&=2f^2\left(1+z+2\mu\right), \label{mR20} \\
m_I^2(0)&=2f^2\left(1+z-2\mu\right).
\end{align}
Our scenario requires that the potential is stable in $\phi_R,\vphi$ and unstable in $\rho$ around $\rho=0$. For this, we impose that $\mu>0$ and $m_I(0)^2>0$, that is,
\begin{align}\label{mu}
\mu>0 \quad \mbox{ and } \quad 1+z-2\mu>0.
\end{align}
In accordance with this condition, we introduce a parameter $\vep$ by
\begin{align}
\label{muzvep}
\vep^2=1+z-2\mu.
\end{align}
The mass function $m_I^2$ at each non-vanishing $\rho$ behaves as
\begin{align}
m_I^2(\rho)=2f^2\vep^2-2f^2(4Az+4\ga+4A\vep^2-\vep^4+f^{-2}q^2z)\,\rho^2+\cO(\rho^4).
\end{align}
As $\rho$ increases from zero, this is positive for a while, but becomes zero at 
\begin{align}\label{rhoc}
\rho_c^2\simeq\frac{\vep^2}{4Az+4\ga+4A\vep^2-\vep^4+f^{-2}q^2z}.
\end{align}
For consistency, we need to require $\rho_c<1$ and thus $\epsilon<1$ too. After $\rho_c$, $\vphi$ becomes tachyonic and the inflaton will likely fall into this direction (``waterfall'') due to quantum fluctuations. This requires a dedicated careful investigation that goes beyond the scope of our paper.

\subsection{Inflation}
We assume that inflation ends by $\rho_c$, so that it is of single-field, hilltop type and described by the potential at $\phi=0$, around the origin of $\rho$. This is in the framework of inflation by supersymmetry breaking~\cite{Antoniadis:2017gjr,Antoniadis:2018cpq,Antoniadis:2019dpm,Aldabergenov:2021uye}, as already described in the Introduction.
We summarise some results that will be used later. 

The inflaton potential $V_{\mathrm{inf}}=V(\rho,\phi=0)$ is given by
\begin{align}
V_{\mathrm{inf}}(\rho)=f^2e^{\rho^2+A\rho^4+B\rho^6}\left[\tfrac{(1+\rho^2+2A\rho^4+3B\rho^6)^2}{1+4A\rho^2+9B\rho^4}-3\rho^2\right]+\tfrac{1}{2}q^2(1+\rho^2+2A\rho^4+3B\rho^6)^2.
\end{align}
The slow-roll parameters then read
\begin{align}
&\ep_V=\left(\frac{8Af^2-2q^2}{2f^2+q^2}\right)^2\rho^2+\cO(\rho^4), \\
&\eta_V=-\frac{8Af^2-2q^2}{2f^2+q^2}+\cO(\rho^2).
\end{align}
Note that the leading order of $\eta_V$ can be tuned at will, while $\ep_V$ is much smaller, $\ep_V\ll{\cal O}(\eta_V^2)$. This is an important feature of models in the framework of inflation by supersymmetry breaking~\cite{Antoniadis:2017gjr}. 

The amplitude of scalar curvature fluctuations $\cA_s$, tilt $n_s$, and tensor-to-scalar ratio $r$ for the cosmic microwave background (CMB) are given by
\begin{align}
n_s=1-6\ep_V(\rho_*)+2\eta_V(\rho_*)\simeq1+2\eta_V(\rho_*), \quad
r=16\ep_V(\rho_*), \quad
\cA_s=\frac{V_{\mathrm{inf}}(\rho_*)}{24\pi^2\ep_V(\rho_*)},
\end{align}
where $\rho_*$ is $\rho$ at the horizon exit. Combining them, we can express the model parameters $\rho_*,A,f$ in terms of the CMB parameters $n_s,r,\cA_s$ as
\begin{align}\label{inf-quantities}
\rho_*&\simeq\frac{\sqrt{r}}{2(1-n_s)}, \quad
A\simeq\frac{1-n_s}{8}+\frac{(5-n_s)q^2}{8(3\pi^2r\cA_s-q^2)}, \quad
f^2\simeq\frac{3\pi^2r\cA_s-q^2}{2}.
\end{align}
In the limit where $U(1)_R$ D-term contributions are neglected when the corresponding gauge coupling is very small (as will be justified in Section~\ref{subsec:addD}), $q$ can be set to zero, leading essentially to one parameter $A$ fixed by the spectral index $n_s$, while the inflation scale $f$ is constrained by the observational data according to \eqref{inf-quantities} for $q=0$. The horizon exit $\rho_*$ and the end of inflation $\rho_{\mathrm{end}}$ are related via the number of e-folds $\De N$ by
\begin{align}\label{endhe}
\frac{\rho_{\mathrm{end}}}{\rho_*}\simeq\exp\left(\frac{1-n_s}{2}\De N\right).
\end{align}
As mentioned above, $\rho_{\mathrm{end}}$ is required to satisfy $\rho_{\mathrm{end}}\leq\rho_c$. In particular, when the equality holds, $\rho_c$ is determined by $\sqrt r$ through the expression for $\rho_*$ in \eqref{inf-quantities}.

\section{Vacuum structure}

The next task is to find a non-supersymmetric vacuum at $(\rho_m,\phi_m)$. We require that the vacuum energy $V_m=V(\rho_m,\phi_m)$ can be tuned to zero. As we have seen above, the inflaton falls in the direction of the imaginary part $\vphi$, keeping $\phi_R=0$. We may therefore restrict our analysis at $\phi_R=0$. 

Recall that inflation with the subsequent waterfall is supposed to occur for small $\rho$. We can then deal with the potential perturbatively in $\rho$. The strategy of our analysis is that we first find the extrema of the potential at $\rho=0$ as the unperturbed part, and then study how they evolve when $\rho$ is turned on. In the following analysis, we shall use a new parameter $\de$ instead of $\lm$, defined by\footnote{
  Note that $\de^2$ is supposed to be positive, so that the potential at $\rho=0$ is stable in $\phi_R$ around $\phi_R=0$.
}
\begin{align}\label{de2}
\de^2=f(\mu^2-4\lm),
\end{align}
and also consider it to be small. Indeed, as we will see below, for $\delta=0$ one obtains a supersymmetric vacuum with a flat direction along $\rho$; thus, the parameter $\delta$ is expected to control the scale of supersymmetry breaking. We will therefore analyse the potential perturbatively both in $\rho$ and in $\de$.

\subsection{Potential at $\rho=0$}\label{subsec:Vrho0}
Let us first consider the potential at $\rho=0$. Since the D-term potential at $\rho=0$ is just a constant in $\phi$, it is sufficient to analyse the F-term potential. The F-term potential at $\rho=0$ reads
\begin{align}
V_F(0,\phi)=\frac{e^{|\phi|^2}|w(\phi)|^2}{1-z\,|\phi|^2}.
\end{align}
Its extrema with zero or small $\phi_R$ are given by
\begin{align}
0, \quad \phi_1^-, \quad \phi_s^\pm, \quad \phi_0, \quad \phi_1^+. \nn
\end{align}
They are given explicitly by
\begin{align}
&\phi_s^\pm=i\vphi_s^\pm
=i\sqrt{\frac{2}{\mu\pm f^{-1/2}\de}}
=i\left(\sqrt{\frac{2}{\mu}}\mp\frac{\de}{\sqrt{2f\mu^3}}\right)+\cO(\de^2), \\
&\phi_1^\pm=i\vphi_1^\pm
=i\left[\frac{\mu-3z\mu+2z\pm\sqrt{(\mu+z\mu-2z)(\mu+9z\mu-2z)}}{2z\mu}\right]^{\frac{1}{2}}+\cO(\de^2), \\
&\phi_0=i\vphi_0
=i\left[\sqrt{\frac{2}{\mu}}+\de^2\frac{\mu-2z+2\mu^2-3z\mu}{2\sqrt{2}f\mu^{7/2}(\mu-2z)}\right]+\cO(\de^4).
\label{phi0-de2}
\end{align}
Notice that they are all pure imaginary.

\begin{figure}
\centering
  \includegraphics[width=\linewidth]{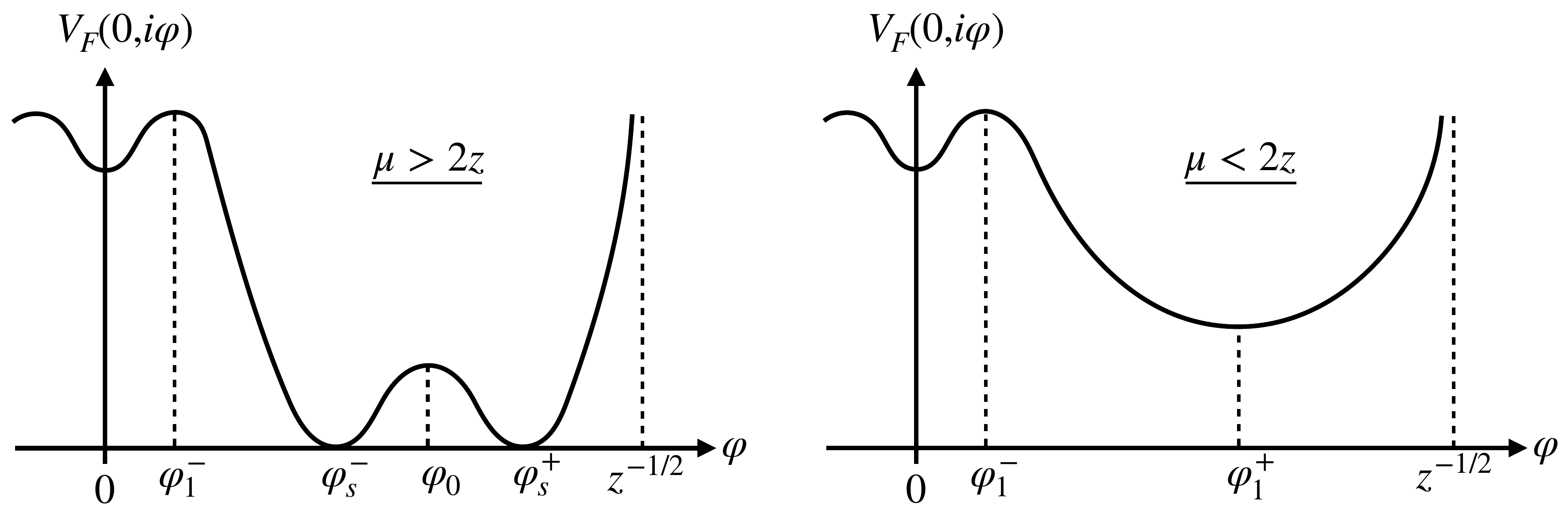}
\caption{Structure of the F-term potential with $\rho=0$ is given for the cases $\mu>2z$ and $\mu<2z$. 
The potential is stable (minimum) at $\phi_R=0$ in both cases. Only $\vphi$-axis (the imaginary axis) is shown. The region $\vphi>z^{-1/2}$ is not drawn because the K\"ahler metric becomes negative.}\label{VFrho0}
\end{figure}

The potential is singular for $|\phi|=z^{-1/2}$.
Its denominator is the K\"ahler metric $K_{X\bar X}$ for $\rho=0$,
which is positive if $|\phi|<z^{-1/2}$. The extrema with $|\phi|<z^{-1/2}$ differ depending on whether the parameters $\mu,z$ satisfy $\mu>2z$ or $\mu<2z$, which are given by:
\begin{align}\label{extremacases}
\begin{split}
\mu>2z\quad \left(z<\frac{1-\varepsilon^2}{3}\right)&: \quad 0, \quad \phi_1^-, \quad \phi_s^\pm, \quad \phi_0, \\
\mu<2z\quad \left(z>\frac{1-\varepsilon^2}{3}\right) &: \quad 0, \quad \phi_1^-, \quad \phi_1^+\,,
\end{split}
\end{align}
where $\varepsilon$ is defined in \eqref{muzvep}.
Whether $\mu>2z$ or $\mu<2z$, the potential is stable in $\phi_R$. On the other hand, the stability around each extremum of \eqref{extremacases} in $\vphi$ is summarised in the following table:
\begin{center}
\begin{tabular}{|c|c|}
\hline
\multicolumn{2}{|c|}{$\mu>2z$} \\
\hline
\hline
$\phi=0$ & min in $\vphi$ \\
\hline
$\phi=\phi_1^-$ & max in $\vphi$ \\
\hline
$\phi=\phi_s^\pm$ & min in $\vphi$  \\
\hline
$\phi=\phi_0$ & max in $\vphi$ \\
\hline
\end{tabular}
\qquad
\begin{tabular}{|c|c|}
\hline
\multicolumn{2}{|c|}{$\mu<2z$} \\
\hline
\hline
$\phi=0$ & min in $\vphi$ \\
\hline
$\phi=\phi_1^-$ & max in $\vphi$ \\
\hline
$\phi=\phi_1^+$ & min in $\vphi$ \\
\hline
\end{tabular}
\end{center}
The potential for each case is depicted schematically in Figure~\ref{VFrho0}.

\paragraph{Case $\boldsymbol{\mu>2z}$:}
The extremum $\phi=0$ is a local minimum in both $\phi_R$ and $\vphi$, as already stated in Section~\ref{subsec:Model}.
Next, $\phi_1^-$ is a saddle point near the origin. Indeed, $\phi_1^-$ is of order $\vep$,
\begin{align}
&\phi_1^-=i\frac{2\vep}{\sqrt{(1-z)(1+3z)}}+\cO(\vep^2,\de^2).
\end{align}
Next, $\phi_s^\pm$ are supersymmetric minima as the superpotential, $D_XW$ and $D_\phi W$ vanish.
Finally, the extremum $\phi_0$ is a saddle point that is located near (sandwiched by) $\phi_s^\pm$.
The value of the F-term potential at $\phi=\phi_0$ is
\begin{align}
V_F(0,\phi_0)&=\frac{e^{2/\mu}}{\mu^3(\mu-2z)}\de^4+\cO(\de^6).
\end{align}
The stability around $\phi_0$ is dictated by
\begin{align}
&\left.\frac{\pd^2V_F}{\pd\phi_R^2}\right|_{\rho=0,\phi=\phi_0}
=\frac{8e^{2/\mu}f}{\mu-2z}\de^2+\cdots, \quad
\left.\frac{\pd^2V_F}{\pd\vphi^2}\right|_{\rho=0,\phi=\phi_0}
=-\frac{8e^{2/\mu}f}{\mu-2z}\de^2+\cdots, \\
&\left.\frac{\pd^2V_F}{\pd\vphi\pd\phi_R}\right|_{\rho=0,\phi=\phi_0}=0.
\end{align}

\paragraph{Case $\boldsymbol{\mu<2z}$:}
The extrema $\phi=0,\phi_1^-$ have the same properties as in the case $\mu>2z$. The extremum $\phi_1^+$ is a minimum, which becomes
\begin{align}
&\phi_1^+=i\sqrt{\frac{(1-z)(1+3z)}{z(1+z)}}+\cO(\vep^2,\de^2),
\end{align}
where we replaced $\mu$ by $\vep^2$ through \eqref{muzvep}. The value of the F-term potential at $\phi_1^+$ is given by
\begin{align}\label{VFrho=0vphi0mu<2z}
V_F(0,\phi_1^+)&=f^2e^{\frac{(1+3z)(1-z)}{z(1+z)}}\frac{(z+1)^5(3z-1)^3}{256z^5}+\cO(\vep^2,\de^2).
\end{align}
Since $1+z-2\mu=\vep^2$ and $2z>\mu>0$, we have $z>1/3+\cO(\vep^2)$ and hence the minimum energy is positive.

\subsection{Potential at nonzero $\rho$}

Before we proceed to detailed analysis of the potential at finite $\rho$, we give a brief summary on the structure of the potential. In this paper, we only consider the case $\mu>2z$. The other case $\mu<2z$ will be briefly summarised later. Its main drawback is that the vacuum energy cannot be tuned to zero consistently with our perturbative treatment since its leading order value \eqref{VFrho=0vphi0mu<2z} is of order unity.

\begin{figure}
\centering
  \includegraphics[width=0.9\linewidth]{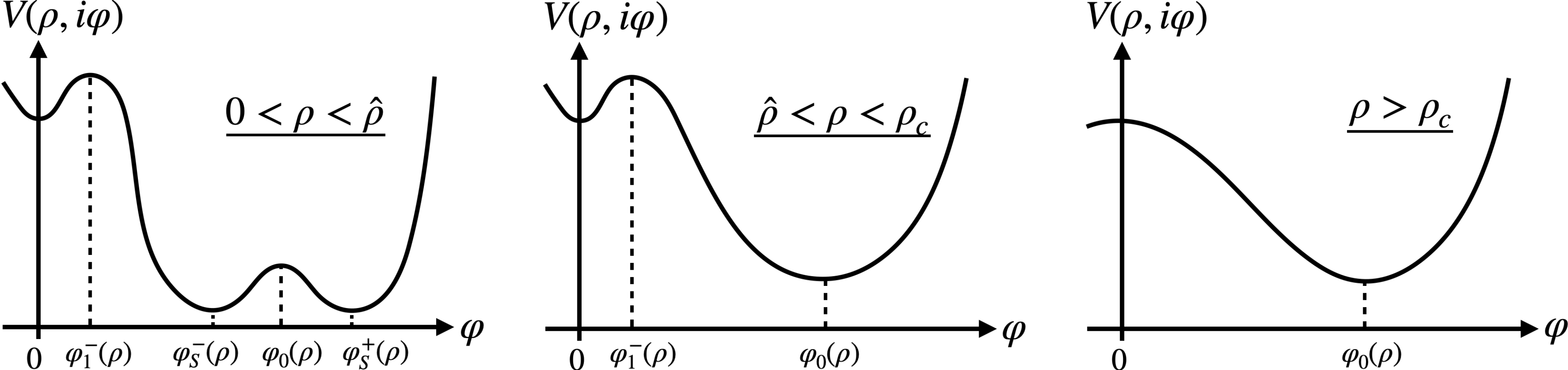}
\caption{Potential at $\rho$ slices in the case $\mu>2z$.}\label{VFrho}
\end{figure}

As $\rho$ is turned on perturbatively, the potential will keep for a while the same structure of the extrema, which we denote by $0,\phi_1^-(\rho),\phi_0(\rho),\phi_s^\pm(\rho)$. 
The potential continues to have the extrema $0,\phi_1^-(\rho)$ at values of order of the inflaton potential energy, while the other extrema at $\phi_0(\rho),\phi_s^\pm(\rho)$ have much smaller energy of order $\de^4$. Remarkably, as we shall see soon below, when $\rho$ becomes greater than some small value $\hat\rho$, the three points $\phi_0(\rho),\phi_s^\pm(\rho)$ will merge into a \textit{single minimum} (being pure imaginary) with energy of order $\de^4$. In other words, at $\hat\rho$, the extremum $\phi_0(\rho)$ changes from a ridge to a valley.
Recall also that $0$ and $\phi_1^-(\rho)$ merge at $\rho_c$ to form the critical point as explained in Section~\ref{subsec:Model}. Thus, the potential at each $\rho>\hat\rho$ slice has only one maximum at $\phi=0$ and one minimum at $\phi_0(\rho)$. In particular, the line $\{\phi_0(\rho):\rho\geq\hat\rho\}$ forms a valley and the minimum of the potential along this valley $V(\rho,\phi_0(\rho))$ with respect to $\rho$ is the true minimum, namely the vacuum $(\rho_m,\phi_0(\rho_m))$.
 Figure~\ref{VFrho} illustrates the potential at three $\rho$-slices.

As mentioned above and will be shown later, the potential along the line $\{\phi_0(\rho):\rho\geq0\}$ is of order $\de^4$, while when $\de=0$, the line $\{\phi_0(\rho):\rho\geq0\}$ reduces to the supersymmetric flat direction $\phi(\rho)=i\sqrt{2/\mu}$. This indicates that $\de^2$ controls the supersymmetry breaking scale, which is true as will be demonstrated in Section~\ref{sec:susybreaking}. Since the inflation scale does not involve $\de^2$, the supersymmetry breaking scale is not correlated with the inflation scale. It will also be shown in Section~\ref{subsec:tuning} that the tuning of the vacuum energy does not affect the inflation scale. 

\paragraph{Summary on the case $\boldsymbol{\mu<2z}$.}
As $\rho$ is turned on perturbatively, the potential will keep the same structure of the extrema, which we denote by $0,\phi_1^\pm(\rho)$. The behaviour of $0,\phi_1^-(\rho)$ is the same as in the case $\mu>2z$. On the other hand, $\phi_1^+(\rho)$ continues to be a minimum (in other words, the potential forms a valley along $\phi_1^+(\rho)$ for $\rho\geq 0$). Therefore, as in the case $\mu>2z$, the inflaton falls in the $\vphi$ direction and reach the minimum at some $\phi_1^+(\rho_m)$ which gives the true minimum among $\phi_1^+(\rho)$'s for $\rho\geq 0$.

However, the vacuum energy $V(\rho,\phi_1^+(\rho_m))$ is independent of $\de^2$, as already indicated in \eqref{VFrho=0vphi0mu<2z}. This is a consequence of the fact that in the $\mu<2z$ case, the potential cannot have the supersymmetric flat direction when $\de=0$, in contrast with the case $\mu>2z$. The vacuum energy and the supersymmetry breaking scales are both controlled by $\mu$ and $z$, and also the supersymmetry breaking scales are correlated with the inflation scales in general. We will not therefore consider this case further.

\subsubsection{Analysis in the absense of the D-term potential}

We first analyse the potential in the case without its D-term part. The D-term contribution will be taken into account in such a manner that it does not destroy the structure of the potential obtained by the analysis without the D-term contribution.

$\phi_0(\rho)$ is defined by the extremality condition
\begin{align}
\label{dVI1}
0=\left.\frac{\pd V_F(\rho,\phi)}{\pd\vphi}\right|_{\phi=\phi_0(\rho)}.
\end{align}
Let us consider the perturbative expansion of $\phi_0$ in $\de$ and $\rho$. First, $\phi_0$ depends on $\de,\rho$ only through $\de^2,\rho^2$ because so does the potential. Second, when $\de=0$, $\phi_0(\rho)$ becomes $i\sqrt{2/\mu}$, which is constant in $\rho$. As a result, the expansion in $\de$ and $\rho$ must have the form
\begin{align}\label{phi0-exp}
\phi_0(\rho)=i\vphi_{(0,0)}+i\sum_{m \geq 1, n \geq 0}\de^{2m}\rho^{2n}\vphi_{(m,n)}.
\end{align}
According to $\phi_0(0)=\phi_0$ given by \eqref{phi0-de2}, the coefficients $\vphi_{(0,0)},\vphi_{(1,0)}$ are given by
\begin{align}
\vphi_{(0,0)}&=\sqrt{\frac{2}{\mu}}, \quad
\vphi_{(1,0)}=\frac{\mu-2z+2\mu^2-3z\mu}{2\sqrt{2}f\mu^{7/2}(\mu-2z)}. \label{phi0-(00)(10)}
\end{align}

Next, let us consider the structure of the potential along $\phi_0(\rho)$. As we explained above, the goal is to find a nontrivial minimum of the potential along the valley of $\vphi$-minima $V_F(\rho,\phi_0(\rho))$ with respect to $\rho$ in a perturbative manner. We therefore need its perturbative expansion in $\rho$ at least up to $\rho^4$. It turns out that the leading term of the expansion in $\de$ is of order $\de^4$,
\begin{align}
V_F(\rho,\phi_0(\rho))=\de^4\frac{e^{2/\mu}}{2\mu^4}\sum_{k \geq 0}v_{Fk}\rho^{2k}+\cO(\de^6),
\end{align}
and the coefficients $v_{F0},v_{F1},v_{F2}$ are determined only by $\vphi_{(0,0)},\vphi_{(1,0)}$, being independent of the other coefficients $\vphi_{(m,n)}$. The three coefficients are given by
\begin{align}\label{vF024}
\begin{split}
&v_{F0}=\frac{2\mu}{\mu-2z}, \quad
v_{F2}=-\frac{16\mu\ga'}{(\mu-2z)^2},  \\
&v_{F4}=\frac{1}{\mu(\mu-2z)^3}\Big[-18B(\mu-2z)\mu^3+\mu^4-4(2z+7\ga')\mu^3 \\
&\qquad\qquad\qquad\qquad~+8(3z^2+14z\ga'+16\ga'^2)\mu^2-16(2z+7\ga')z^2\mu+16z^4\Big], 
\end{split}
\end{align}
where we introduced $\ga'=\ga+(A\mu/2)$ to absorb $A$.\footnote{
This can be understood as follows: at the leading order in $\de$, the $X^2\bar X^2$ term in the K\"ahler potential becomes $(A+2\ga/\mu)X^2\bar X^2$, so $A$ appears in $V_F(\rho,\phi_0(\rho))$ up to $\rho^4$ only through the combination $A+2\ga/\mu$.}
The stability at the extremum $\phi_0(\rho)$ can be seen by expanding the potential there:
\begin{align}
\left.\frac{\pd^2V_F}{\pd\phi_R^2}\right|_{\phi=\phi_0(\rho)}
&=\frac{8e^{2/\mu}f}{\mu-2z}\de^2+32e^{2/\mu}f^2\mu^2\rho^2+\cdots, \label{VFstability1} \\
\left.\frac{\pd^2V_F}{\pd\vphi^2}\right|_{\phi=\phi_0(\rho)}
&=-\frac{8e^{2/\mu}f}{\mu-2z}\de^2+32e^{2/\mu}f^2\mu^2\rho^2+\cdots. \label{VFstability2}
\end{align}
They imply that $\phi=\phi_0(\rho)$ is a minimum if $\rho$ satisfies (as long as $\rho$ is smaller than 1)
\begin{align}\label{rhohat}
\rho^2>\hat\rho^2:=\frac{1}{4f\mu^2(\mu-2z)}\de^2.
\end{align}

\subsubsection{Adding the D-term potential}\label{subsec:addD}
Let us turn on the D-term potential. We regard it as a perturbative correction so that the vacuum structure we found for the F-term potential should not get destroyed. Since $V_F(\rho,\phi_0(\rho))$ is of order $\de^4$, a natural choice of $q$ may be $q\sim\de^2$, so that $V_D\propto q^2\sim\de^4$. We therefore set
\begin{align}\label{qhat}
q=e^{1/\mu}\mu^{-2}\hat q\de^2.
\end{align}

We can expand $\phi_0(\rho)$ in $\de,\rho$ in the same way as in the case without the D-term. Moreover, the D-term potential vanishes at $\rho=0$. Therefore, the expansion of $\phi_0(\rho)$ has the form \eqref{phi0-exp} with the coefficients $\vphi_{(0,0)},\vphi_{(1,0)}$ given by \eqref{phi0-(00)(10)}. The value of the potential at $\phi_0(\rho)$ for each $\rho$ at the leading order in $\de$ is still of order $\de^4$ due to our choice of $q$ above:
\begin{align}
V(\rho,\phi_0(\rho))=\de^4\frac{e^{2/\mu}}{2\mu^4}\sum_{k \geq 0}v_k\rho^{2k}+\cO(\de^6),
\end{align}
and the coefficients $v_0,v_2,v_4$ depend only on $\vphi_{(0,0)},\vphi_{(1,0)}$, independent of the other coefficients. The three coefficients are given by
\begin{align}\label{v024}
\begin{split}
&v_0=v_{F0}+\hat q^2, \quad
v_2=v_{F2}+\frac{2(\mu-2z)}{\mu}\hat q^2, \\
&v_4=v_{F4}+\frac{\mu^2-4(z-2\ga')\mu+4z^2}{\mu^2}\hat q^2,
\end{split}
\end{align}
where $v_{F0},v_{F2},v_{F4}$ are given in \eqref{vF024}.
We can also show that the D-term potential does not affect the stability at $\phi_0(\rho)$ given by \eqref{VFstability1} and \eqref{VFstability2} for each $\rho$. Therefore, $\hat\rho$ is the same as \eqref{rhohat}, and $\phi_0(\rho)$ is the minimum for each $\rho>\hat\rho$.

The vacuum is obtained by minimising $V(\rho,\phi_0(\rho))$ with respect to $\rho$ in the region $\hat\rho<\rho<1$,
\begin{align}
&V_m=V(\rho_m,\phi_0(\rho_m))=\de^4\frac{e^{2/\mu}}{2\mu^4}\left(v_0-\frac{v_2^2}{4v_4}\right), \quad
\rho_m^2=-\frac{v_2}{2v_4}. \label{Vm-rhom}
\end{align}
Note that $v_0>0$.
For $V_m$ to be a minimum, we need
\begin{align}
\label{min-conds}
v_4>0, \qquad
v_2<0.
\end{align}
Let us come back to the consistency conditions for our model. For our scenario to work, $\hat\rho<\rho_c<\rho_m$ is needed. Combining this with \eqref{rhoc<rhom<1} and taking the perturbative nature of the analysis above into account, we require
\begin{align}\label{rhohat<<rhoc<rhom<1}
\hat\rho<\rho_c<\rho_m<1.
\end{align}

\subsection{Tuning of the vacuum energy}
\label{subsec:tuning}

Let us require that the vacuum energy $V_m$ is zero, $V_m=0$, which yields $v_2^2=4v_0v_4$. Since $v_0>0$, the positivity of $v_4$ is automatically guaranteed. Therefore, we only have to impose $v_2<0$ out of the conditions in \eqref{min-conds}. Remarkably, the equation $v_2^2=4v_0v_4$ is just a first-order polynomial in $B$, so exactly solvable for $B$. Let us denote the solution by $B=B_m$, which is a quadratic polynomial in $\ga'$,
\begin{align}\label{Bquad}
\begin{split}
B_m&=B_{m0}+B_{m1}\ga'+B_{m2}\ga'^2,
\end{split}
\end{align}
where the coefficients are given by
\begin{align}\label{Bcoeff}
\begin{split}
B_{m0}&=\frac{(\mu-2z)^3(2\mu+3\hat q^2(\mu-2z))}{18\mu^3(2\mu+\hat q^2(\mu-2z))}, \\
B_{m1}&=-\frac{2(\mu-2z)(14\mu^2-\hat q^2\mu(\mu-2z)-2\hat q^4(\mu-2z)^2)}{9\mu^3(2\mu+\hat q^2(\mu-2z))}, \\
B_{m2}&=\frac{32(3\mu+2\hat q^2(\mu-2z))}{9\mu(\mu-2z)(2\mu+\hat q^2(\mu-2z))}.
\end{split}
\end{align}
Under this choice of $B$, the $\rho$-coordinate $\rho_m$ at the Minkowski vacuum reads
\begin{align}\label{rhomink}
\rho_m^2=-\frac{2v_0}{v_2}=\frac{\mu(\mu-2z)(2\mu+\hat q^2(\mu-2z))}{8\mu^2\ga'-\hat q^2(\mu-2z)^3}.
\end{align}
Its positivity is guaranteed by the negativity of $v_2$ in \eqref{min-conds}.

To check the validity of the model around the vacuum, we need the trace and determinant of the K\"ahler metric $g_{i\bar j}=K_{i\bar j}$ at the vacuum. For generic $\rho,B$, they read
\begin{align}
\tr\, g&=1+\det g+\cO(\rho^4), \\
\det g&=\frac{\mu-2z}{\mu}+\left(\frac{8\ga'}{\mu}-z\right)\rho^2+\cO(\rho^4),
\end{align}
where $g_{i\bar j}=K_{i\bar j}$. The trace is automatically positive if the determinant is positive.

\subsection{Positivity of mass squared}
The $U(1)_R$ gauge boson is massive around the Minkowski vacuum by Higgs mechanism. When the model is coupled with MSSM, the sparticles in the MSSM sector acquire soft masses. We require that they are not tachyonic. 

The mass squared of the $U(1)_R$ gauge boson around the vacuum is given by
\begin{align}
m_A^2&=2q^2K_{X\bar X}\rho_m^2, \label{vecboson}
\end{align}
where the K\"ahler metric component $K_{X\bar X}$ satisfies
\begin{align}\label{KXX-detg}
K_{X\bar X}-\det g=[z\rho^2+\cO(\rho^4)]+\cO(\de^2).
\end{align}
The positivity of $K_{X\bar X}$ is guaranteed as long as $\det g>0$ and $z>0$.

Next, let us consider soft scalar masses. Note that the coupling with the MSSM sector does not change the vacuum energy because the vacuum expectation value of each MSSM field vanishes at the vacuum. Let $\zt$ collectively denote squarks and sleptons. The soft scalar mass squared of $\zt$ is given by
\begin{align}
m_0^2(Q_\zt)=\bra V_F \ket+\bra e^KW\ol W \ket+q^2Q_\zt\bra 1+XK_X \ket, \label{m0(Q)2}
\end{align}
where $Q_\zt=0$ in Model 1 and $Q_\zt=1/2$ in Model 2 for every $\zt$, and the bracket $\bra\cdots\ket$ means the vacuum expectation value. This mass squared is not positive definite in general because $V_F$ must be negative at the Minkowski vacuum, $\bra V_F \ket=-\bra V_D \ket<0$. Its explicit expression is given by
\begin{align}\label{m0(Q)2-explicit}
m_0^2(Q_\zt)=\de^4\frac{e^{2/\mu}}{\mu^4}\left[\rho_m^2-\hat q^2\left\{\frac{1}{2}-Q_\zt+(1-Q_\zt)\frac{\mu-2z}{\mu}\rho_m^2\right\}+\cO(\rho_m^4)\right]+\cO(\de^6).
\end{align}
The positivity condition on $m_0^2(Q_\zt)$ in each case is given by
\begin{align}
\mbox{Case 1:} &\quad m_0^2(0)>0, \\
\mbox{Case 2:} &\quad m_0^2(1/2)>0.
\end{align}

\section{Allowed parameter region}
\label{sec:constraints}

Our model depends on the following parameters $A,B,z,\ga,f,\mu,\de,\hat q$. We have already fixed $B$ by tuning the vacuum energy to zero. We will find the allowed region for the other parameters by imposing the observational and theoretical constraints.

The observational constraints come from the consistency of the inflation period of our model with the observational data on CMB~\cite{Planck:2018jri}. This yields constraints on the model parameters through the formulas \eqref{inf-quantities}. We will use the following values~\cite{Planck:2018jri}:
\begin{align}\label{cmb-vals}
n_s\simeq0.96, \quad \cA_s\simeq2.1\times10^{-9}, \quad \De N\simeq60.
\end{align}
Actually, the number of e-folds $\De N$ is constrained from below by a value around 34.2 plus a logarithmic contribution depending on reheating temperature~\cite{Planck:2018jri}, 
$\Delta N\gtrsim 34.2+\ln\left(T_\text{reh}/1\,\text{TeV}\right)$,
allowing for a wider range of $\De N$ than our choice above. However, in our setup $\De N$ appears only through the factor $e^{(1-n_s)\De N/2}$ (see eq.\eqref{endhe}). Even if we adopt for example $\De N=40\sim60$, this factor varies only from 2.2 to 3.3, which does not substantially affect the allowed parameter region we find below.

We have also imposed various conditions to guarantee the consistency of our analysis. They are summarised as follows:
\begin{align}
&|\phi_0| \gg \de^2|\vphi_{(1,0)}|, \label{phim} \\
&\hat\rho<\rho_c < \rho_m < 1, \label{rhos} \\
&\rho_{\mathrm{end}} \leq \rho_c \label{endc} \\
&A,B_m,\ga<1, \label{pert} \\
&v_2<0, \quad v_4>0, \label{v2v4min} \\
&\mbox{no ghost around the vacuum}, \label{trdet} \\
&m_0^2(Q)>0 \quad (Q=0 \mbox{ for Model 1}, ~ Q=1/2 \mbox{ for Model 2}). \label{m0Q2>0}
\end{align}
Their meanings are summarised as follows:
The first condition is needed to justify the perturbative expansion in $\de^2$.
The second one \eqref{rhos} is nothing but \eqref{rhohat<<rhoc<rhom<1}.
The third one \eqref{endc} means that the inflation is of the single-field type until its end.
The fourth one \eqref{pert} requires our model should be in a perturbative regime as an effective theory. 
The fifth one \eqref{v2v4min} guarantees that the vacuum is indeed a minimum.
The sixth one \eqref{trdet} requires the theory around the vacuum to be physical. This condition consists of two parts: one for the kinetic terms of the inflaton and waterfall field, $\det g_{i\bar j}>0$, and the other for the kinetic term of the fermion orthogonal to the goldstino, which will be given later in \eqref{fermion_kin}. 
The seventh one \eqref{m0Q2>0} means that the soft scalar masses for quarks and leptons are not tachyonic.

The analysis of the constraints in what follows consists of two parts: one for upper bounds on $\de^2$ and the other to find an allowed region for the other parameters. To facilitate the analysis, we introduce $x$ by
\begin{align}
x=\frac{\mu-2z}{\mu}.
\end{align}
We also introduced $\vep^2=1+z-2\mu$ \eqref{muzvep}. Combining these gives
\begin{align}
\mu=\frac{2(1-\vep^2)}{3+x}, \quad
z=\frac{(1-x)(1-\vep^2)}{3+x}.
\end{align}
We will use $x,\vep^2$ instead of $\mu,z$. Since $x,\vep^2>0$, we have $\mu<2/3$ and $z<1/3$. Moreover, $\mu>0$ yields $\vep^2<1$.

\subsection{Upper bounds on $\de^2$ and tensor-to-scalar ratio}
 We first present upper bounds on $\de^2$. Technical details are given in Appendix~\ref{app:parameter_space}. The constraints \eqref{phim} and $\hat\rho<\rho_c$ in \eqref{rhos} yield the following upper bound on $\de^2/\sqrt r$:
\begin{align}\label{de2-pertvphi}
\frac{\de^2}{\sqrt{r}} \ll  \max\left\{5.4\times10^{-5}(1-\vep^2)^3,~2.1\times10^{-4}\frac{\vep^2x}{4\ga-\vep^4}\right\}.
\end{align}
On the other hand, the following upper bound
\begin{align}\label{de2-simple}
\frac{\de^2}{\sqrt{r}} \ll \frac{2.4\times10^{-6}}{\hat q}
\end{align}
allows us to neglect the D-term contributions $q^2$ in the expressions \eqref{inf-quantities} for $A$ and $f$:
\begin{align}\label{f2simple}
A\simeq 0.005, \qquad f^2 \simeq \frac{3}{2}\pi^2r\cA_s.
\end{align}
We will choose $\de^2/\sqrt r$ so that the two upper bounds are satisfied. As will be demonstrated in the next section, $\de^2$ controls the supersymmetry breaking scale. Therefore, the upper bound on $\de^2$ gives a strong suppression of the supersymmetry breaking scales and soft masses.

Since the horizon exit $\rho_*$ is proportional to $\sqrt r$ according to \eqref{inf-quantities},
we can find an upper bound on $r$ from the condition $\rho_{\mathrm{end}}^2<1$ from \eqref{rhos},
\begin{align}
r<\frac{4(1-n_s)^2}{e^{(1-n_s)\De N}}\rho_m^2\sim5.8\times10^{-4}\rho_m^2<5.8\times10^{-4}.
\end{align}

\subsection{Allowed parameter region for $(x,\ga')$}

Let us analyse the other conditions: concretely,
\begin{align}\label{conds_others}
\rho_{\mathrm{end}}\leq\rho_c<\rho_m<1, \quad
\det g_{i\bar j}>0, \quad k_{\phi\bar\phi}>0, \quad
m_0^2(Q)>0, \quad B_m<1, \quad \ga<1,
\end{align}
where we dropped $A<1$ because it has already been achieved in \eqref{f2simple}, and $k_{\phi\bar\phi}$ is the coefficient of the kinetic term of the fermion orthogonal to the goldstino, which will be given by \eqref{fermion_kin} in the next section. We will find the allowed region in the $(x,\ga')$-plane, regarding $\vep,\hat q,r$ as parameters put by hand. The difference from the last analysis for $\de^2$ is that the analysis here is independent of $\de^2$. In this section, we will just summarise the result, delegating technical details to Appendix~\ref{app:parameter_space}. 

For each $x$, the conditions $\rho_{\mathrm{end}}\leq\rho_c$ and $m_0(Q)^2>0$ bound $\ga'$ from above, while $\rho_c<\rho_m<1$ and the conditions for the correct signs of the kinetic terms bound $\ga'$ from below. In particular, in Model 1, compatibility of the upper bound on $\ga'$ from $m_0(Q)^2>0$ with the value of $\ga'$ for a given $\rho_m^2$ yields an upper bound on $\hat q^2$,
\begin{align}\label{hatq-ubound-model1}
\hat q^2<2\rho_m^2.
\end{align}
Since $\rho_m^2$ should be substantially smaller than 1 for the validity of the perturbative analysis, it is natural to suppose $\hat q<1$. On the other hand, in Model 2, $m_0(Q)^2>0$ does not yield an upper bound on $\ga'$ but just $x\hat q^2<2$. Therefore, Model 2 does not have a counterpart of the condition \eqref{hatq-ubound-model1}.

\begin{figure}
\begin{tabular}{cc}
\begin{minipage}[t]{\textwidth}
  \centering 
  \includegraphics[width=0.8\linewidth]{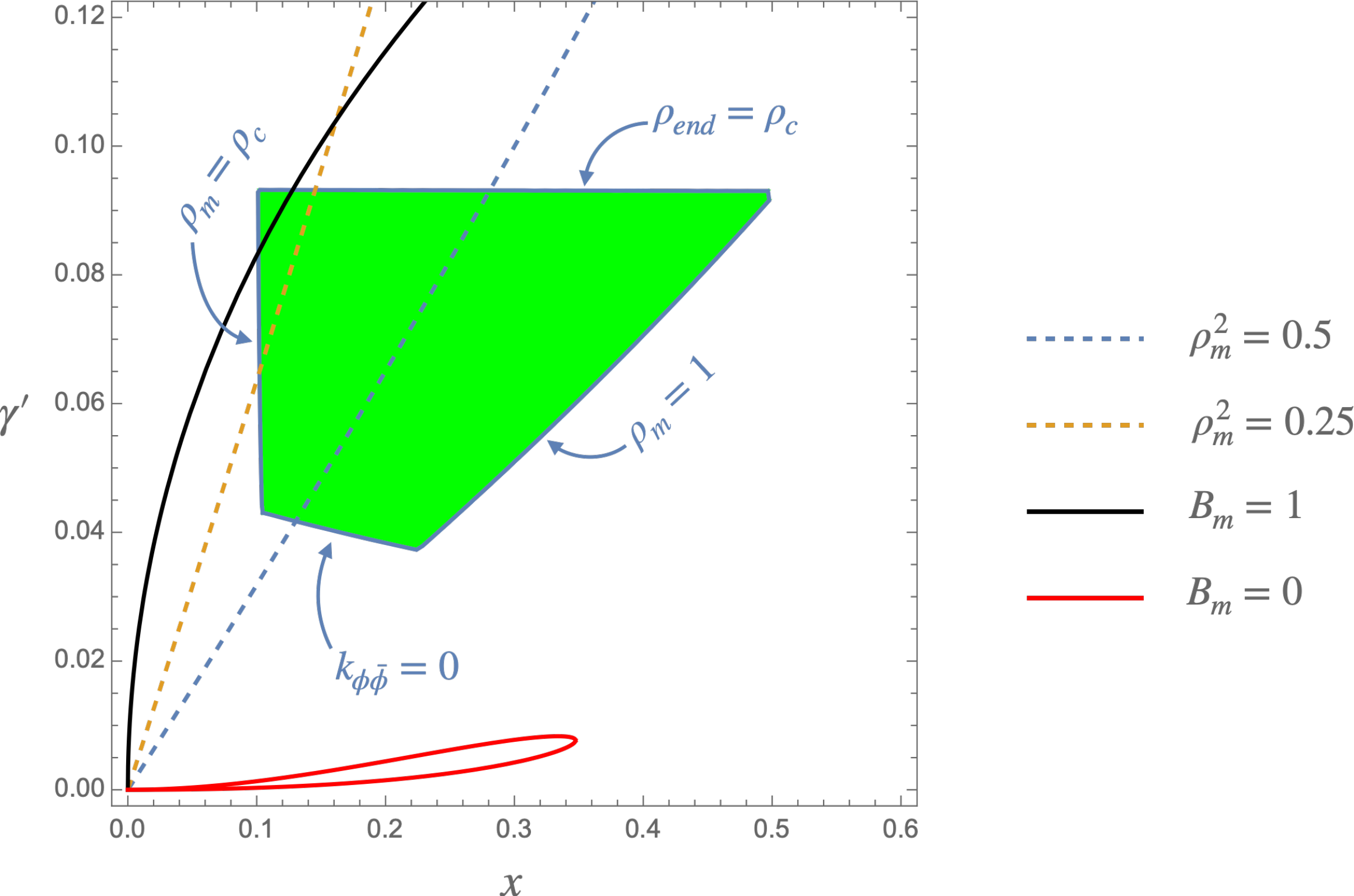}
  \caption{The green region shows the allowed parameter region for $(\hat q,\vep)=(1,0.25)$. The tensor-to-scalar ratio is $r=1.0\times10^{-4}$. Model 1 and Model 2 have the same allowed region.}\label{x_gamma1}
\end{minipage} \\
\mbox{} \\
\mbox{} \\
\begin{minipage}[t]{\textwidth}
  \centering 
  \includegraphics[width=0.8\linewidth]{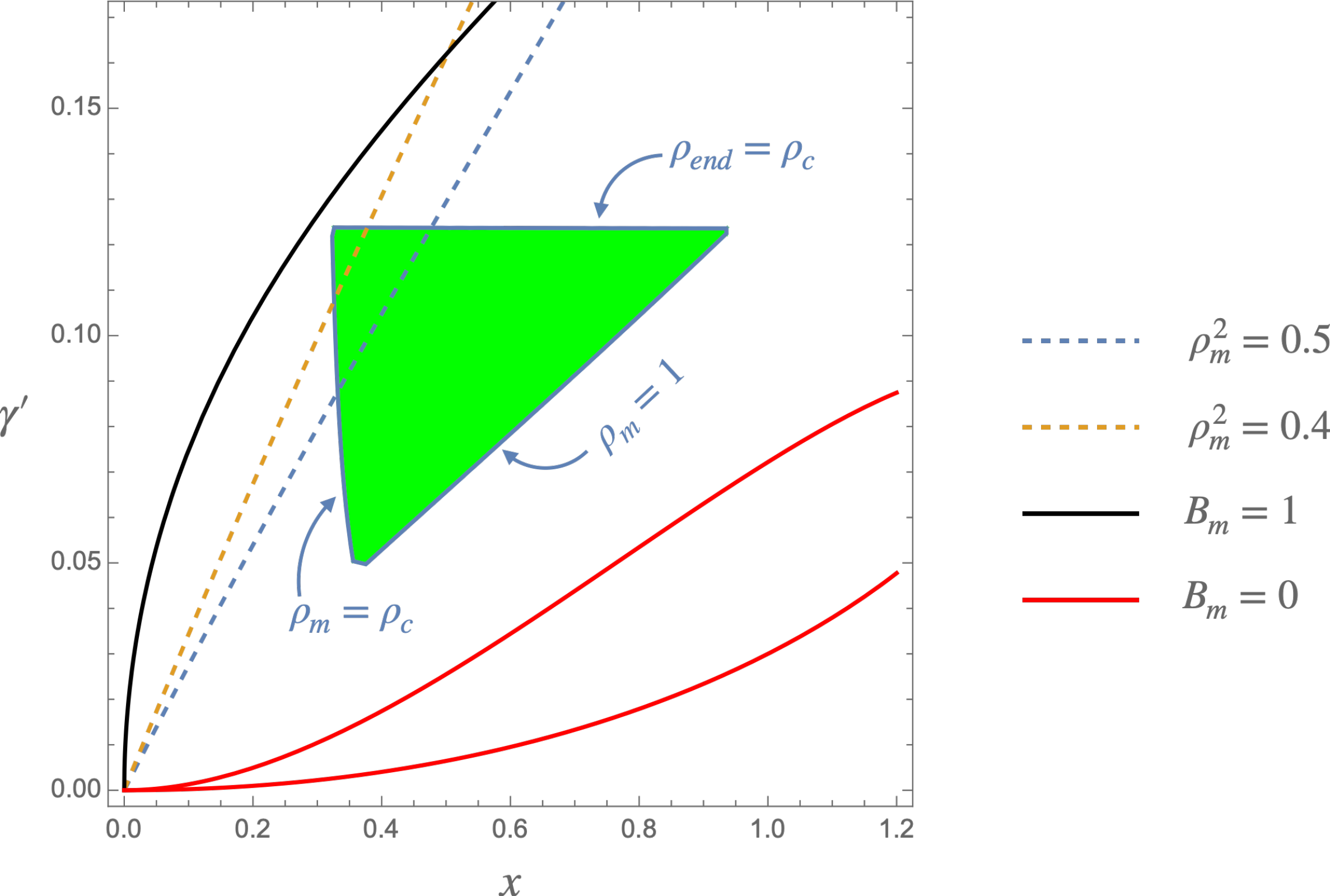}
  \caption{The green region shows the allowed parameter regions for $(\hat q,\vep)=(0.5,0.4)$. The tensor-to-scalar ratio is $r=2.0\times10^{-4}$. Model 1 and Model 2 have the same allowed region.}\label{x_gamma2}
\end{minipage}
\end{tabular}
\end{figure}

Here we give examples: Figure~\ref{x_gamma1} shows the allowed region for $(\hat q,\vep,r)=(1,0.25,1.0\times10^{-4})$ in Model 1, and Figure~\ref{x_gamma2} shows the allowed region for $(\hat q,\vep,r)=(0.5,0.4,2.0\times10^{-4})$ in Model 1.
There the conditions \eqref{pert} are satisfied on the right of the $B_m=1$ line. The condition $m_0(Q)^2>0$ does not contribute to the boundary of the allowed regions in these examples because it is weaker than the other conditions. This is true also in Model 2, so the allowed region for Model 2 for each case is the same. 

As seen from the figures, the allowed region is upper bounded by the curve for $\rho_{\mathrm{end}}=\rho_c$, which is almost constant in $x$ because according to the second inequality of \eqref{rhoc>=rhoend1}, its $x$-dependence is $A/(3+x)$ with $A$ being as small as $0.005$. Its value is given by
\begin{align}\label{rhoc>=rhoend}
\ga'\lesssim\frac{(1-n_s)^2}{e^{(1-n_s)\De N}}\frac{\vep^2}{r}+\frac{\vep^4}{4}
\sim\vep^2\left(\frac{1.5\times10^{-4}}{r}+\frac{\vep^2}{4}\right),
\end{align}
where we simplified the expression \eqref{rhoc} for $\rho_c^2$ by neglecting the terms other than $4\ga,\vep^4$, which is true as long as $\vep<1$ is chosen as in the examples above.\footnote{The strict upper bound on $\de^2$ allows neglecting $f^{-2}q^2z$.} We can see that for a given $\vep$, this upper bound on $\ga'$ is fixed by the tensor-to-scalar ratio. In particular, when $x$ is chosen from the upper bound line $\rho_{\mathrm{end}}=\rho_c$, the tensor-to-scalar ratio fixes $\ga'$. 

On the other hand, since the curve for $\rho_m=1$ increases monotonically, it will meet the line for $\rho_{\mathrm{end}}=\rho_c$ at some $x$, which gives the maximum value for $x$, which is given approximately as
\begin{align}
x\lesssim\frac{3\vep^2(4(1-n_s)^2+e^{(1-n_s)\De N}\vep^2r)}{(2-2\vep^2-\vep^4)r-4\vep^2(1-n_s)^2},
\end{align}
where we used $\hat q^2x, \hat q^2x^2\ll 2$. In this approximation, the positivity of its right hand side, which comes from $x>0$, gives a lower bound on $r$,
\begin{align}
r>\frac{4\vep^2(1-n_s)^2e^{(1-n_s)\De N}}{2-2\vep^2-\vep^4}.
\end{align}

Here we make comments on possibilities of reducing parameters. We already pointed out that under the identification $\rho_{\mathrm{end}}=\rho_c$, the tensor-to-scalar ratio fixes the values of $\ga'$. Note that $\ga'=0$ is excluded because $x>0$ by definition and hence $\rho_m^2<1$ forces $\ga'>0$. Moreover, $\ga=0$, namely $\ga'=A\mu/2$, is also excluded: the condition $k_{\phi\bar\phi}>0$ forces $\vep$ to be very close to $1$, which, however, makes $\vep^4$ dominant in the denominator of $\rho_c^2$ in \eqref{rhoc} and hence $\rho_c^2<0$. On the other hand, one may wonder the possibility of $B_m=0$, but it is also impossible, as demonstrated in the figures: The condition $B_m=0$ yields two roots that form two curves in the $(x,\ga')$-plane, and they do not intersect with the allowed region.

\section{Supersymmetry breaking scale}
\label{sec:susybreaking}

In this section, we discuss supersymmetry breaking around the vacuum and the particle spectrum in both hidden and observable sectors.

\subsection{Supersymmetry breaking scale}
The gravitino mass is given by
\begin{align}\label{m32}
|m_{3/2}|=|\bra\ka^2e^{\ka^2K}W\ol W\ket|^{1/2}
\simeq\frac{e^{1/\mu}}{\mu^2}\rho_m\de^2,
\end{align}
where the bracket $\bra\cdots\ket$ denotes the vacuum expectation value. The non-vanishing vacuum expectation values of the auxiliary fields in the chiral supermultiplets $X,\phi$ and in the R-vector multiplet are given by
\begin{align}
&|\bra F^X \ket|=|\bra e^{K/2}g^{X\bar i}D_{\bar i}\ol W\ket|
\simeq\frac{e^{1/\mu}}{\mu(\mu-2z)}\de^2, \\
&|\bra F^\phi \ket|=|\bra e^{K/2}g^{\phi\bar i}D_{\bar i}\ol W \ket|
\simeq\frac{\sqrt{2}(\mu-2z+2\mu z)}{\mu^{5/2}(\mu-2z)}\rho_m\de^2, \\
&|\bra \cD \ket|=|\bra q(1+XK_X) \ket|\simeq\frac{e^{1/\mu}}{\mu^2}\hat q\de^2.
\end{align}
The ratio $|\bra F^\phi \ket/\bra F^X \ket|$ is of order $\rho_m$, and the ratio $|\bra D \ket|/|\bra F^X \ket|$ is of order $x\hat q$. Therefore, both ratios are smaller than 1 as long as $x$ and $\hat q$ are taken to be smaller than 1 and supersymmetry breaking is dominated by the F-auxiliary of the inflaton multiplet. Thus, its fermionic component is mainly the goldstino that makes it superparnter of the inflaton, as expected in the framework of ``inflation by supersymmetry breaking''. It is also manifest from the last equations that $\de^2$ controls the supersymmetry breaking scale. 

\subsection{Particle spectrum}

\subsubsection*{Bosonic spectrum}
The bosonic particles in the hidden (inflaton) sector are two scalars and one $U(1)_R$ vector boson. The mass of the vector boson $m_A$ and the masses $m_w,m_i$ of the two scalars at the vacuum are given by
\begin{align}
m_A\simeq \frac{e^{1/\mu}}{\mu^2}\sqrt{\frac{2(\mu-2z)}{\mu}}\rho_m\hat q\de^2, \quad
m_i\simeq \frac{e^{1/\mu}}{\mu}\sqrt{\frac{-2v_2}{\mu(\mu-2z)}}\de^2, \quad
m_w\simeq 4fe^{1/\mu}\mu\rho_m,
\end{align}
where $m_A$ comes from \eqref{vecboson}. The expression inside the square root of $m_i$ is positive because $v_2<0$ \eqref{min-conds}. The mass $m_w$ corresponds to a scalar linear combination dominated mainly by the waterfall field $\vphi$, while $m_i$ corresponds to a linear combination dominated mainly by the inflaton $\rho$. The real part of the waterfall field $\phi_R$ has the mass~$\simeq4fe^{1/\mu}\mu\rho_m$ equal to $m_w$ up to $\cO(\de^2)$.

In the observable MSSM sector, as we have already discussed, the sfermions acquire a common mass $m_0(Q)$ given by \eqref{m0(Q)2-explicit}, where $Q=0$ in Model 1 and $Q=1/2$ in Model 2.

\subsubsection*{Fermionic spectrum}
The fermion mass spectrum can be read off from the action in the super-unitary gauge given in \eqref{fermion_action_sunitary}. In our setup, the coefficients in the kinetic terms are given by
\begin{align}\label{fermion_kin}
k_{\phi\bar\phi} \simeq 1-\frac{z}{x}\rho_m^2, \quad 
k_{RR} \simeq 1+\frac{1}{2}\hat q^2x, \quad 
k_{\phi R} = \cO(\hat q\de^2), \quad
k_{i\bar j}=\de_{i\bar j}, \quad
k_{aa} = 1+\bt_a\ln\rho_m,
\end{align}
and the other coefficients are zero. The subscript `R' refers to $U(1)_R$, and the index $a$ refers to the MSSM gauge groups: $U(1),SU(2),SU(3)$. The indices $i,\bar j$ refer to the MSSM chiral superfields.\footnote{Note the difference from the definition in Appendix~\ref{app:fmass}, where the indices $i,\bar j$ refer not only to the MSSM chiral superfields but also to the waterfall.} The coefficients $\bt_a$, together with $\bt_R$ for $U(1)_R$ in \eqref{gk}, are determined by the cancellation of the anomalies involving $U(1)_R$. In our models they read~\cite{Aldabergenov:2021uye}
\begin{align}
\mbox{Model 1:} &\quad
\bt_R=-\frac{q^2}{3\pi^2}, \quad
\bt_{U(1)}=-\frac{11g_1^2}{8\pi^2}, \quad
\bt_{SU(2)}=-\frac{5g_2^2}{8\pi^2}, \quad
\bt_{SU(3)}=-\frac{3g_3^2}{8\pi^2}, \\
\mbox{Model 2:} &\quad
\bt_R=-\frac{7q^2}{48\pi^2}, \quad
\bt_{U(1)}=\frac{g_1^2}{8\pi^2}, \quad
\bt_{SU(2)}=\frac{g_2^2}{8\pi^2}, \quad
\bt_{SU(3)}=\frac{3g_3^2}{8\pi^2}, \quad
\end{align}
where $g_1,g_2,g_3$ are the gauge coupling constants for $U(1),SU(2),SU(3)$, respectively. 
 
The fermions in the hidden sector are the $U(1)_R$ gaugino and the fermion orthogonal to the goldstino which is eaten by the gravitino. Their masses are given by
\begin{align}
m_{\lm_R}\simeq\frac{(\mu-2z)^2+4\mu\ga'}{\mu^3(\mu-2z)}\rho_m\hat q^2\de^2, \quad
m_f\simeq 4fe^{1/\mu}\mu\rho_m, 
\end{align}
where $m_f$ is equal to $m_w$ up to $\cO(\de^2)$. In the supersymmetric limit $\de=0$, the fermion mass $m_f$ and the waterfall boson mass $m_w$ become exactly equal, leading to a massive waterfall supermultiplet.

The fermions in the observable sector are the higgsinos and the gauginos for the MSSM gauge groups. The higgsinos have mass $\mu\rho_m$. The gaugino mass $m_a$ for the MSSM gauge group labelled by $a$ is given by
\begin{align}
m_a\simeq\frac{e^{1/\mu}}{2\mu(\mu-2z)}\frac{\bt_a}{\rho_m}\de^2.
\end{align}
Notice that we can control the overall scale of $m_a,m_i,m_0,m_{\lm_R}$ by tuning $\de^2$. All particles with masses proportional to $\de^2$ constitute the light spectrum of the theory. Besides the MSSM states, these are the inflaton, the R-gauge boson, the gravitino and the R-gaugino.

Since the light states charged under $U(1)_R$ have comparable mass and charge of order $\de^2$ in Planck units, it is interesting to check the weak gravity conjecture postulating that gravity is the weakest force in Nature and black-holes decay without remnants~\cite{Arkani-Hamed:2006emk}. This requires the existence of at least one state whose $U(1)$ charge is no less than its mass. In our case, for Model 1, the lightest charged state is the gravitino which has $U(1)_R$ charge $q/2$ and mass $m_{3/2}$, given by \eqref{qhat} and \eqref{m32}. Imposing the condition $q/2\ge m_{3/2}$ one finds
\begin{align}\label{wgc}
\hat q \geq 2\rho_m.
\end{align}
It turns out that in Model 1, this lower bound contradicts the upper bound \eqref{hatq-ubound-model1}. On the contrary, Model 2 survives because $\hat q$ is just subject to the bound $\hat q^2<2x^{-1}$, which, combined with the lower bound \eqref{wgc}, yields $x<(2\rho_m^2)^{-1}$. On the other hand in Model 2, MSSM fields with odd R-parity are also charged under $U(1)_R$ and in particular electroweak gauginos can be lighter than the gravitino (see the example below), which weakens the constraint \eqref{wgc}.

We finish our analysis by presenting as benchmark model an explicit example with low mass spectrum in the multi-TeV region within the reach of next generation of particle colliders. It contains a neutralino, mostly electroweak gaugino, as the lightest supersymmetric particle (LSP) which is a standard Dark Matter candidate and predicts the existence of the universal inflaton sector ($U(1)_R$ gauge boson and gaugino, as well as the inflaton) among the low energy spectrum, which is the smoking gun of inflation by supersymmetry breaking. 
Let us consider Model 2 in Figure~\ref{x_gamma1}, where the choice $(\hat q,\rho_m^2)=(1,1/4)$ satisfies the lower bound \eqref{wgc} from the weak gravity conjecture applied to the gravitino. We choose $(x,\ga')\simeq(0.1454,0.09318)$, which is the intersection of the $\rho_m^2=1/4$ and $\rho_{\mathrm{end}}=\rho_c$ lines. The constraint on $\de^2$ is $\de^2\ll2.4\times10^{-8}$. For example, we may choose $\de^2$ so that the masses of the soft scalars and MSSM gauginos are around $10\sim100$\,TeV. If we choose $\de^2\sim10^{-14}$ and use the values for the coupling constants of the MSSM gauge groups (see for instance~\cite{Aldabergenov:2021uye}),
\begin{align}\label{mssm_gauge_couplings}
g_{U(1)}\sim0.5, \quad g_{SU(2)}\sim0.59, \quad g_{SU(3)}\sim0.72,
\end{align}
the lightest super-particle is the $U(1)$ gaugino (bino) of mass $\sim16$\,TeV. The masses of the other particles are (in TeV)
\begin{alignat}{5}
&m_i \simeq 7.3\times10^3, &\quad~
&m_A \simeq 97, &\quad~
&m_w \simeq 2.6\times10^{10}, &\quad~
&m_0 \simeq 170, &\quad~ &\\
&m_{3/2} \simeq 180, &\quad~
&m_{\lm_R} \simeq 150, &\quad~
&m_f \simeq 2.6\times10^{10}, &\quad~
&m_{SU(2)} \sim 22, &\quad~
&m_{SU(3)} \sim 98.
\end{alignat}
Notice that the waterfall field is not part of the light spectrum; it is superheavy, of the order of the inflation scale.

The formula for the soft scalar masses $m_0^2$ is valid only for the squarks and sleptons, while the masses for the Higgs fields get an additional contribution from the superesymmetric mass term $\mu_HH_uH_d$ in the superpotential \eqref{model2-W} that introduces an additional mass parameter $\mu_H$. It is convenient for the following to introduce $\hat\mu_H\equiv \mu_H/m_{3/2}$ which satisfies: 
\begin{align}
\hat\mu_H=\frac{\mu_H}{fw(\phi_m)} \quad;\quad \phi_m\equiv \phi_0(\rho_m)\,,
\end{align}
where $\phi_m$ is the vacuum expectation value of the waterfall field at the supersymmetry breaking minimum of the scalar potential~\eqref{Vm-rhom}. The Higgs masses and mixing are then given by:
\begin{align}
m_{H_u} \simeq m_{H_d}\simeq m_{3/2}(1+\hat\mu_H^2)^{1/2} \quad;\quad
B\mu_H=\hat B\hat\mu_Hm_{3/2}^2\quad;\quad \hat B\simeq 10\,,
\end{align}
where we kept only the dominant contribution, neglecting higher order corrections in the K\"ahler potential and corrections from $U(1)_R$ gauge interactions. Obviously, these contributions are important when $\mu_H$ is large, i.e. $\hat\mu_H$ of order unity, which may be required in some models for the avoidance of the cosmological moduli problem (see e.g.~\cite{Bae:2022okh}).

\section{Conclusion}

In this work we presented an extension of the simple framework called `inflation from supersymmetry breaking' by introducing a waterfall field, allowing to stop inflation fast and to decouple parametrically the supersymmetry breaking and inflation scales. The idea behind inflation from supersymmetry breaking consists of identifying the inflaton with the superpartner of the goldstino in the presence of a gauged R-symmetry and the inflation sector with the supersymmetry breaking sector in the context of gravity mediation models. The gauge R-symmetry fixes the inflaton superpotential, absorbs its pseudo-scalar companion into the massive R-gauge boson and makes slow-roll parameters naturally small, determined by a quartic correction to the canonical inflaton K\"ahler potential. Thus, the minimal field content of the inflaton sector consists of two superfields: one chiral (the inflaton) and one vector (the R-gauge boson). 

The simplest model has then 4 parameters: the inflation scale $f$, the quadratic $A$ and quartic $B$ corrections to the inflaton K\"ahler metric, as well as the $U(1)_R$ gauge coupling normalised at the inflaton charge $q$ which can be chosen sufficiently small so that it plays no role during inflation.
For positive and small $A$, the scalar potential has a maximum at the origin of the inflaton field $\rho$ with a plateau, realising hilltop inflation. The second slow-roll parameter $\eta$ is determined by $A$, while the first slow-roll parameter $\epsilon$ is much smaller: $\epsilon\sim\eta^2\rho_*^2<\eta^2$ with $\rho_*$ the inflaton value at the horizon exit, suppressing the fraction of primordial gravity waves. The value of $A$ is then fixed by the spectral index of primordial scalar perturbations. 
At the end of inflation, the inflaton rolls down to the minimum of the potential with a tuneable infinitesimally small value using the parameter $B$. A microscopic supergravity theory generating the above model and its parameters was provided by a generalisation of the Fayet-Iliopoulos model to a $U(1)_R$~\cite{Antoniadis:2019dpm}. As mentioned above the coupling of the inflaton sector to the supersymmetric Standard Model is straightforward with an interesting possibility of having the usual R-parity as a remnant of the underlying gauged R-symmetry. Moreover, supersymmetry breaking is transmitted from the inflaton to the observable sector as in gravity mediation models, predicting however a heavy spectrum of superparticles, near the inflation scale.

The extension of the above model to hybrid inflation consists in introducing a new chiral superfield $\phi$ (the waterfall) neutral under R-symmetry with a superpotential $w(\phi)$ which is a small deformation of a superpotential $w_0(\phi)$ that admits a supersymmetric vacuum: $\partial_\phi w_0=w_0=0$ at $\phi=\phi_0$. One then writes $w=w_0+\delta^2w_1$ with $\delta$ the deformation parameter that breaks supersymmetry in the vacuum at a scale $\delta^2$ which is different and can be hierarchically smaller than the inflation scale $f$. For simplicity, we considered as an example a quartic function invariant under $\phi\to -\phi$ depending on two couplings that can be traded as $\delta^2$ and a mass $\mu$. The latter controls the critical inflaton value $\rho_c$ at which the waterfall direction opens up because a vanishing expectation value becomes unstable (tachyonic). We impose that this happens after the end of inflation (of at least 60 e-folds). We expect that our results are independent of the particular form chosen for the superpotential. Finally, in analogy with the inflaton K\"ahler metric, we introduced quadratic and quartic (in the inflaton) corrections to the waterfall K\"ahler metric with corresponding parameters $z$ and $\gamma$; these are free parameters of the model, subject to theoretical and experimental constraints.

The requirement of perturbative treatment of both inflation and potential minimum, together with detailed investigation of the parameter space done in Section~4, leads a light spectrum of order $\delta^2 \ll f$ that contains besides the usual MSSM fields (scalar quarks and leptons and SM gauginos) the universal inflaton sector of inflation by supersymmetry breaking: $U(1)_R$ gauge boson and gaugino together with the inflaton. The light mass spectrum has the following approximate pattern: the electroweak gauginos tend to be the lightest as long as shown in the explicit example we presented in Section~5. The next lightest group contains the squarks and sleptons, $U(1)_R$ gauge boson and gaugino, and the gravitino, having masses of more or less the same order. Then the inflaton comes next a bit heavier. This is a model-independent feature of our framework which connects supersymmetry breaking and inflation in a minimal way. Its embedding in a fundamental theory, such as string theory, remains an open challenging question.

\section*{Acknowledgement}
This work was supported in part by the NSRF via the Program Management Unit for Human Resources \& Institutional Development, Research and Innovation [Grant No.~B13F670063]. IA is supported by the Second Century Fund (C2F), Chulalongkorn University.

\appendix

\section{Technical details on the analysis of parameter space}
\label{app:parameter_space}

The condition \eqref{phim} is satisfied if $\de^2/\sqrt{r}$ satisfies the following upper bound:
\begin{align}
\de^2 \ll \frac{32fx(1-\vep^2)^3}{(3+x)^2(x^2+6x+1)}\lesssim5.4\times10^{-5}(1-\vep^2)^3\sqrt{r},
\end{align}
where the numerical value comes from the maximum of the function $x/((3+x)^2(x^2+6x+1))$ at $x\simeq0.4$.

\subsubsection*{Simplification of $A,f$ in \eqref{inf-quantities}}
The conditions to neglect the D-term contributions from the expressions \eqref{inf-quantities} of $A,f$ are given by
\begin{align}
\frac{1-n_s}{8}\ll\frac{(5-n_s)q^2}{8(3\pi^2r\cA_s-q^2)}, \quad
q^2 \ll 3\pi^2r\cA_s.
\end{align}
They give an upper bound on $\de^2/\sqrt{r}$,
\begin{align}\label{app:de2-simple}
\frac{\de^2}{\sqrt{r}} 
\ll \frac{e^{-1/\mu}\mu^2\pi}{\hat q}\sqrt{\frac{3\cA_s(1-n_s)}{2(3-n_s)}}
\sim \frac{2.4\times10^{-6}}{\hat q}.
\end{align}

\subsubsection*{Constraints on quantities at the Minkowski vacuum}
Since $\rho_m^2<1$ is stronger than $v_2<0$, it is enough to impose $\rho_m^2<1$ only. Solving $\rho_m^2=-v_2/(2v_4)$ (see \eqref{Vm-rhom}) for $\ga'$ gives
\begin{align}\label{rhom^2-ga'}
\ga'=\frac{x(\rho_m^{-2}(2+\hat q^2x)+\hat q^2x^2)}{4(3+x)}(1-\vep^2).
\end{align}
The condition $\rho_m^2<1$ gives a lower bound:
\begin{align}\label{rhom^2<1}
\ga'>\frac{x(2+\hat q^2x+\hat q^2x^2)}{4(3+x)}(1-\vep^2).
\end{align}
The positivity of the soft scalar masses squared $m_0(Q)^2>0$ gives 
\begin{align}\label{app:m0(Q)^2>0}
\rho_m^2>\frac{\hat q^2(1-2Q)}{2(1-\hat q^2x(1-Q))} ~~ \mbox{ and } ~~
1-\hat q^2x(1-Q)>0,
\end{align}
where $Q=0$ in Case 1 and $Q=1/2$ in Case 2. 
More concretely, in Model 1, they give the following upper bounds on $\ga'$ and $x$,
\begin{align}\label{ga'soft}
\ga'<\frac{x(4-2\hat q^2x-\hat q^4x^2)}{4\hat q^2(3+x)}(1-\vep^2) ~~ \mbox{ and } ~~
x<\hat q^{-2}.
\end{align}
Compatibility of \eqref{rhom^2-ga'} with \eqref{ga'soft} gives
\begin{align}
0<x<\frac{2-\rho_m^{-2}\hat q^2}{2\hat q^2} ~~\mbox{ and }~~ 
\hat q^2<2,
\end{align}
which in addition yields the condition $\hat q^2<2\rho_m^2$.
On the other hand, in Model 2, the first condition in \eqref{app:m0(Q)^2>0} becomes trivial $\rho_m^2>0$ and hence the only nontrivial constraint is
\begin{align}
x<2\hat q^{-2}. 
\end{align}

The no-ghost condition \eqref{trdet} have two parts: one from the kinetic terms for the inflaton and waterfall field $\det g_{i\bar j}>0$, and the other for the fermion orthogonal to the goldstino $k_{\phi\bar\phi}>0$ given in \eqref{fermion_kin}. The latter one gives the following lower bound on $\ga'$:
\begin{align}
\ga'\geq\frac{(1-x)(1-\vep^2)^2}{(3+x)^2}+\hat q^2\frac{(1-\vep^2)(2(1-x)(1-\vep^2)+x^2(3+x))}{4(3+x)^2}.
\end{align} 
The condition $\det g_{i\bar j}>0$ reads
\begin{align}
x+\left[\frac{4(3+x)\ga'}{1-\vep^2}-\frac{(1-x)(1-\vep^2)}{3+x}\right]\rho_m^2>0.
\end{align}
This is satisfied automatically if the coefficient of $\rho_m^2$ is positive, namely
\begin{align}\label{detg>0-1}
\ga'\geq\frac{(1-x)(1-\vep^2)^2}{4(3+x)^2}.
\end{align}
On the other hand, if this is not satisfied, we must impose
\begin{align}
\ga'<\frac{(1-x)(1-\vep^2)^2}{4(3+x)^2} ~~ \mbox{ and } ~~ \rho_m^2<\frac{x(3+x)(1-\vep^2)}{(1-x)(1-\vep^2)-4(3+x)^2\ga'}.
\end{align}
Compatibility of the first condition of them with \eqref{rhom^2<1} allows a tiny region near $(x,\ga')=(0,0)$. Indeed, we can show that it must be inside the rectangle $0<x\lesssim(\sqrt{57}-7)/4\sim0.137459\cdots$ and $0<\ga'<1/36$. But, as demonstrated in Figures~\ref{x_gamma1} and \ref{x_gamma2}, this region tends to be excluded by the condition $\rho_c<\rho_m$, though not always.

\subsubsection*{Constraints on $\rho^2_c$}

Instead of full analysis, we consider the case where only $4\ga$ and $\vep^4$ are kept in the denominator of the expression of $\rho_c^2$ \eqref{rhoc},
\begin{align}\label{rhoc-simple}
\rho_c^2\simeq\frac{\vep^2}{4\ga-\vep^4}.
\end{align}
This simplification is natural as $A,z$ are as small as $A\sim0.005, z<1/3$ and $f^{-2}q^2z$ is suppressed by $\de^4$ through $q$ (see \eqref{qhat}). Using \eqref{inf-quantities}, we can rewrite the condition $\rho_{\mathrm{end}} \leq \rho_c$ as
\begin{align}
r\leq\frac{4(1-n_s)^2}{e^{(1-n_s)\De N}}\frac{\vep^2}{4\ga-\vep^4} \quad \mbox{ and } \quad 4\ga-\vep^4>0,
\end{align}
which gives the following bounds on $\ga'$ (recall $\ga'=\ga+A\mu/2$),
\begin{align}\label{rhoc>=rhoend1}
\frac{\vep^4}{4}+\frac{A(1-\vep^2)}{3+x}<\ga'\leq\frac{(1-n_s)^2}{e^{(1-n_s)\De N}}\frac{\vep^2}{r}+\frac{\vep^4}{4}+\frac{A(1-\vep^2)}{3+x}.
\end{align}
Since $A\simeq0.005$, the $x$-dependence of the bounds is negligible, so that the upper bound is approximately constant in $x$.
\begin{align}\label{rhoc>=rhoend2}
\ga'\lesssim\frac{(1-n_s)^2}{e^{(1-n_s)\De N}}\frac{\vep^2}{r}+\frac{\vep^4}{4}
\sim\vep^2\left(\frac{1.5\times10^{-4}}{r}+\frac{\vep^2}{4}\right) \quad (\De N=60).
\end{align}
Next, $\rho_c<\rho_m$ gives
\begin{align}
\left[x(2+\hat q^2x)-\frac{\vep^2(3+x)}{1-\vep^2}\right]\ga'>\left(\frac{\vep^4}{4}-\frac{A(1-\vep^2)}{3+x}\right)x(2+\hat q^2x)-\frac{\hat q^2\vep^2x^3}{4}.
\end{align}
The condition $\hat\rho<\rho_c$ from \eqref{rhos} yields an upper bound on $\de^2/\sqrt r$:
\begin{align}\label{rhoc2torhohat2}
\frac{\de^2}{\sqrt{r}}<\frac{32\vep^2(1-\vep^2)^3x}{(4\ga-\vep^4)(3+x)^3}\sqrt{\frac{3\pi^2\cA_s}{2}}\lesssim2.1\times10^{-4}\frac{\vep^2x}{4\ga-\vep^4}.
\end{align}

\section{Fermion masses}
\label{app:fmass}
Let $\Phi^I$ collectively denote the chiral superfields $X,\phi$ and the MSSM ones denoted by $\zt$, and $\chi^I$ be the fermionic component of $\Phi^I$. 
The fermions in our models have the following quadratic Lagrangian: 
\begin{align}
 -\tfrac{1}{2}ig_{I\bar J}\bar\chi^{\bar J}\bar\sig^\mu\pd_\mu\chi^I
 -\tfrac{1}{2}ih_{AB}^*\bar\lm^B\bar\sig^\mu\pd_\mu\lm^A
 -\tfrac{1}{2}m_{IJ}\chi^I\chi^J
 -m_{IA}\lm^A\chi^I
 -\tfrac{1}{2}m_{AB}\lm^A\lm^B
 +\mbox{c.c.}\,,
\end{align}
where the labels $i,j,\cdots$ refer to the chiral superfields except $X$. The labels $A,B,\cdots$ refer to the gauged $U(1)_R$ and the MSSM gauge groups $U(1),SU(2),SU(3)$, which we denote by $G_A$. The labels $a,b,\cdots$ refer to the MSSM gauge groups only, which we denote by $G_a$. The gauge kinetic functions are chosen to be 
\begin{align}
h_{AB}=h_A\de_{AB}, \quad h_A=1+\bt_A\ln X,
\end{align}
where the coefficients $\bt_A$ are determined by the cancellation of the anomalies involving $U(1)_R$. The mass coefficients are given by
\begin{align}
&m_{IJ}=W_{IJ}+K_{IJ}W+K_I\nab_JW+K_J\nab_IW-K_IK_JW-\Ga^K_{IJ}\nab_KW, \\
&m_{AI}=-\sqrt{2}g_{I\bar J}\bar X^{\bar J}_A+\tfrac{\sqrt{2}}{4}i\pd_Ih_{AB}\cD^B, \\
&m_{AB}=\tfrac{1}{2}F^I\pd_Ih_{AB},
\end{align}
where $K_{IJ}=\pd_I\pd_JK$ and $\Ga^K_{IJ}=g^{\bar LK}\pd_Jg_{I\bar L}$.
 $X^I_A$ is the Killing vector for $\Phi^I$, defined by splitting off transformation parameters from the infinitesimal transformation of $\Phi^I$ under gauge group $G_A$. In our setup $m_{aI}=0$ holds since $\phi$ is neutral under $U(1)_R$ and the MSSM fields have vanishing expectation values at the vacuum. The auxiliary fields are defined by
\begin{align}
F^I=-e^{K/2}g^{I\bar J}\nab_J\bar W, \quad
\cD_A=iX_A^I(K_I+W^{-1}W_I), \quad
\cD^A=(h^{-1})^{AB}\cD_B.
\end{align}

In the superunitary gauge, the goldstino $\eta$ is set to zero:
\begin{align}
\eta=-\tfrac{1}{\sqrt{2}}ig_{I\bar J}\chi^I\bar F^{\bar J}+\tfrac{1}{2}\cD_A\lm^A=0.
\end{align}
Eliminating $\chi^X$ from the action by solving this constraint gives
\begin{align}\label{fermion_action_sunitary}
\begin{split}
&-\tfrac{1}{2}ik_{i\bar j}\bar\chi^{\bar j}\bar\sig^\mu\pd_\mu\chi^i
-\tfrac{1}{2}ik_{iA}\bar\lm^A\bar\sig^\mu\pd_\mu\chi^i
-\tfrac{1}{2}ik_{A\bar i}\bar\chi^{\bar i}\bar\sig^\mu\pd_\mu\lm^A
-\tfrac{1}{2}ik_{AB}\bar\lm^B\bar\sig^\mu\pd_\mu\lm^A \\
&-\tfrac{1}{2}\wt{m}_{ij}\chi^i\chi^j
-\wt{m}_{iA}\lm^A\chi^i
-\tfrac{1}{2}\wt{m}_{AB}\lm^B\lm^A \\
&+\mbox{c.c.}\,,
\end{split}
\end{align}
where the coefficients in the kinetic terms are given by
\begin{align}
k_{i\bar j}&=g_{i\bar j}
-g_{X\bar j}\frac{F_i}{F_X}
-g_{i\bar X}\frac{\bar F_{\bar j}}{\bar F_{\bar X}}
+g_{X\bar X}\frac{F_i\bar F_{\bar j}}{F_X\bar F_{\bar X}}, \\
k_{iA}&=\frac{1}{\sqrt{2}}ig_{i\bar X}\frac{\cD_A}{\bar F_{\bar X}}
-\frac{1}{\sqrt{2}}ig_{X\bar X}\frac{\cD_AF_i}{F_X\bar F_{\bar X}}, \\
k_{AB}&=h_{AB}^*+\frac{1}{2}g_{X\bar X}\frac{\cD_A\cD_B}{F_X\bar F_{\bar X}},
\end{align}
and the coefficients in the mass terms are given by
\begin{align}
\wt{m}{}_{ij}&=m_{ij}-m_{Xi}\frac{F_j}{F_X}-m_{Xj}\frac{F_i}{F_X}
+m_{XX}\frac{F_iF_j}{F_X^2}, \\
\wt{m}{}_{iA}&=m_{iA}-m_{XA}\frac{F_i}{F_X}
-\frac{1}{\sqrt{2}}im_{Xi}\frac{\cD_A}{F_X}
+\frac{1}{\sqrt{2}}im_{XX}\frac{\cD_AF_i}{F_X^2}, \\
\wt{m}_{AB}&=m_{\lm\lm}
-\frac{1}{\sqrt{2}}im_{XA}\frac{\cD_B}{F_X}
-\frac{1}{\sqrt{2}}im_{XB}\frac{\cD_A}{F_X}
-m_{XX}\frac{\cD_A\cD_B}{2F_X^2},
\end{align}
where $F_I=g_{I\bar J}\bar F^{\bar J}$.

\bibliography{waterfall.bib}
\bibliographystyle{utphys}

\end{document}